\pdfoutput=1
\documentclass[final,twocolumn,prl,aps,floats,amsfonts,amssymb]{revtex4}
\usepackage{graphicx}
\usepackage{color}
\usepackage{enumerate}
\usepackage{stmaryrd}
\usepackage{amssymb}
\usepackage{amsmath}
\usepackage{hyperref}
\def \degree {^\mathrm{o}}

\begin{document}

\title{Direct measurements of anisotropic energy transfers in a rotating turbulence experiment}

\author{Cyril Lamriben}
\author{Pierre-Philippe Cortet}
\author{Fr\'{e}d\'{e}ric Moisy}
\affiliation{Laboratoire FAST, CNRS, Univ Paris-Sud,
UPMC Univ Paris 06, B\^{a}t. 502, Campus
universitaire, 91405 Orsay, France}

\begin{abstract}
We investigate experimentally the influence of a background
rotation on the energy transfers in decaying grid turbulence. The
anisotropic energy flux density, ${\bf F} ({\bf r}) = \langle
\delta {\bf u}\,(\delta {\bf u})^2 \rangle$, where $\delta {\bf
u}$ is the vector velocity increment over separation ${\bf r}$, is
determined for the first time using Particle Image Velocimetry. We
show that rotation induces an anisotropy of the energy flux
$\nabla \cdot {\bf F}$, which leads to an anisotropy growth of the
energy distribution $E({\bf r}) = \langle (\delta {\bf u})^2
\rangle$, in agreement with the K\'arm\'an-Howarth-Monin equation.
Surprisingly, our results prove that this anisotropy growth is
essentially driven by a nearly radial, but orientation-dependent,
energy flux density ${\bf F} ({\bf r})$.
\end{abstract}

\maketitle

The energy cascade from large to small scales, and the associated
Kolmogorov 4/5th law, are recognized as the most fundamental
results of homogeneous and isotropic
turbulence~\cite{frisch1995,sagaut2008}.  In the presence of a
background rotation, a situation which is relevant for most
geophysical and astrophysical flows, the scale-to-scale energy
transfers are modified by the Coriolis force, yielding a gradual
columnar structuring of turbulence along the rotation
axis~\cite{cambon1989,waleffe1993,staplehurst2008,mininni2010,moisy2011}.
The Taylor-Proudman theorem is often invoked, however improperly,
to justify the resulting quasi-2D nature of turbulence under
rotation. Indeed, this theorem is a purely linear result, which
applies only in the limit of zero Rossby number (i.e. infinite
rotation rate), and is therefore incompatible with turbulence; it
cannot describe the anisotropic energy transfers responsible for
the non-trivial organization of rotating turbulence which are a
subtle non-linear effect taking place only at non-zero Rossby
number. To date, no direct evidence for these anisotropic energy
transfers towards the 2D state in the physical space has been
obtained. In this Letter, we report for the first time direct
measurements of the physical-space energy transfers in decaying
rotating turbulence using Particle Image Velocimetry (PIV), and
provide new insight into the anisotropy growth of turbulence at
finite, and hence geophysically relevant, Rossby number.

\begin{figure}
\centerline{\includegraphics[width=7.5cm]{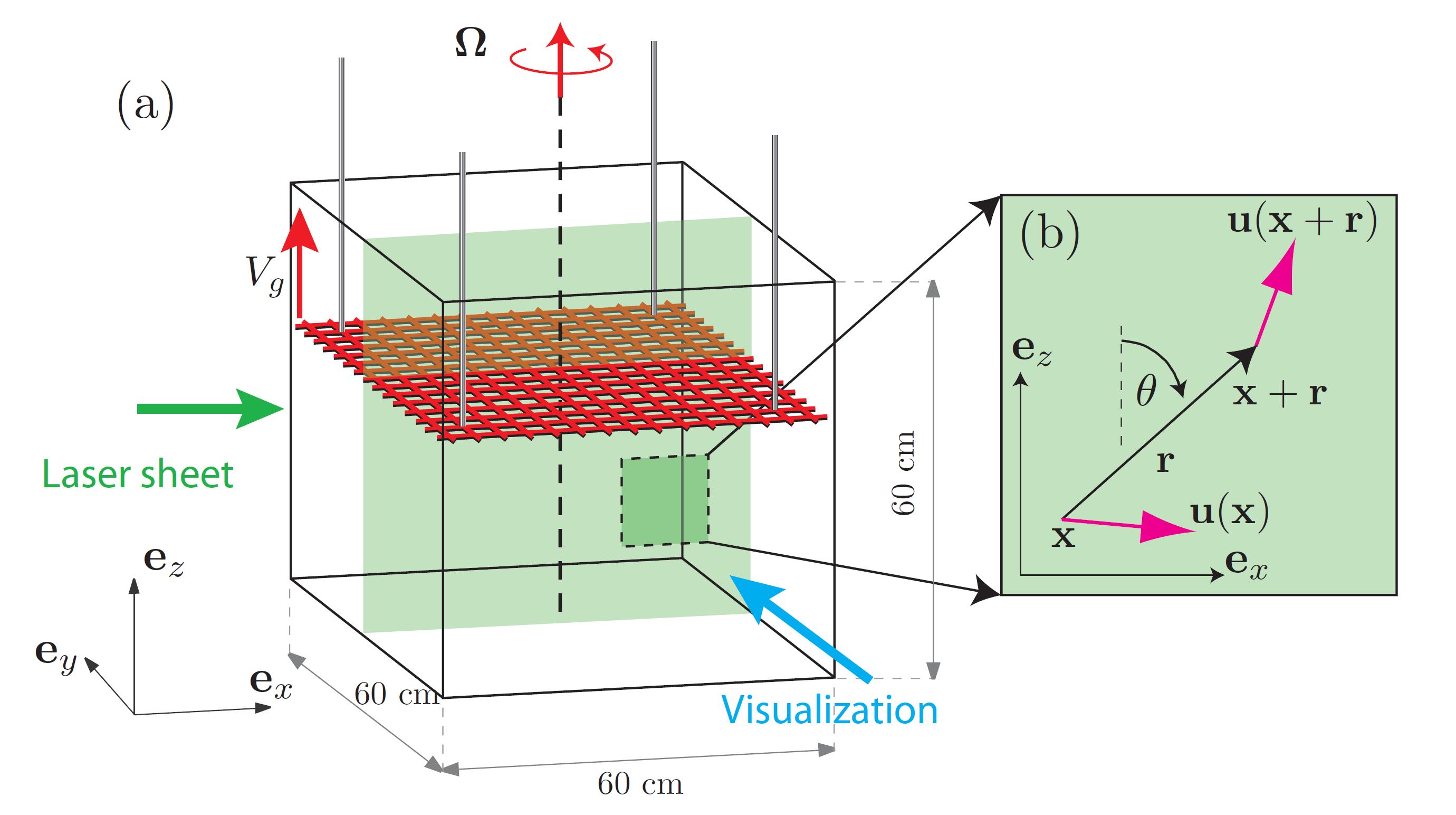}}
\caption{(a) Experimental setup. The water-filled tank is rotating
at $0 \leq \Omega \leq 1.68$~rad~s$^{-1}$. The grid is towed from
the bottom to the top, and PIV measurements are performed in the
vertical plane $(x,z)$ in the rotating frame during the turbulence
decay. (b) Definition of the vector velocity increment $\delta
{\bf u} = {\bf u} ({\bf x}+{\bf r})-{\bf u} ({\bf x})$.}
\label{fig:schemaexp}
\end{figure}

If homogeneity (but not necessarily isotropy) holds, the energy
distribution and energy flux density in the space of separations
${\bf r}$ are described by the fields
\begin{equation}
E ({\bf r},t) = \langle (\delta {\bf u})^2 \rangle \qquad {\rm
and} \qquad {\bf F} ({\bf r},t) = \langle \delta {\bf u}\,(\delta
{\bf u})^2 \rangle, \label{eq:defF}
\end{equation}
where ${\bf u} ({\bf x},t)$ is the turbulent velocity, $\delta
{\bf u} = {\bf u} ({\bf x+r},t) - {\bf u} ({\bf x},t)$ is the
velocity vector increment over ${\bf r}$
(Fig.~\ref{fig:schemaexp}), and $\langle\cdot\rangle$ denotes
spatial and ensemble averages. These key quantities satisfy the
K\'arm\'an-Howarth-Monin (KHM)
equation~\cite{monin1975,frisch1995}, which describes the
evolution of the energy distribution in the space of separations,
\begin{equation}
\frac{1}{2} \frac{\partial}{\partial t} R = \frac{1}{4} \nabla \cdot {\bf F}
+ \nu \nabla^2 R,
\label{eq:vKH}
\end{equation}
where $R({\bf r},t) = \langle {\bf u} ({\bf x},t) \cdot  {\bf u}
({\bf x}+{\bf r},t) \rangle = \langle {\bf u}^2 \rangle - E({\bf
r},t)/2$ is the two-point velocity correlation and $\nu$ the
kinematic viscosity. Importantly, this equation is still valid for
homogeneous anisotropic turbulence~\cite{galtier2009}, and in
particular for axisymmetric turbulence in a rotating frame (here
axisymmetry is to be understood in the statistical sense, with
respect to ${\bf r}$). For stationary (forced) turbulence, this
equation reduces to $\nabla \cdot {\bf F} = -4\,\epsilon$ in the
inertial range, where $\epsilon$ stands for the rates of injected
and dissipated energy. In the isotropic case, this constant-flux
relation yields a purely radial flux density, ${\bf F} ({\bf r}) =
-(4/3)\, \epsilon \,{\bf r}$, describing the usual energy cascade
from large to small scales. This result is actually identical to
the celebrated Kolmogorov's 4/5th law, classically expressed in
terms of the 3rd order longitudinal structure function, $\langle
\delta u_L^3 \rangle = - (4/5)\, \epsilon \, r$, where $\delta u_L
= \delta {\bf u} \cdot {\bf r} / r$ is the longitudinal velocity
increment.

In decaying rotating homogeneous turbulence, Eq.~(\ref{eq:vKH})
shows that, starting from an isotropic initial energy distribution
$E ({\bf r},0)$, an anisotropy growth in $E({\bf r},t)$ is
expected if an anisotropic energy flux $\nabla \cdot {\bf F}$ is
induced by the Coriolis force. However, the flux density ${\bf F}
({\bf r})$ itself has never been measured, and its precise form,
which reveals the fundamental action of rotation on turbulence, is
so far unknown. The only experimental attempts to characterize the
energy transfers in rotating turbulence were restricted to
measurements of $\langle \delta u_L^3 \rangle$ in the plane normal
to the rotation axis~\cite{baroud2002,morize2005}, hence ignoring
the anisotropic nature of those transfers. Recent theoretical
efforts have been made to generalize the 4/5th law, assuming weak
anisotropy~\cite{chakra2007}, or considering the full anisotropic
problem but restricted to the stationary case~\cite{galtier2009}.

\paragraph*{Experiments.-}
The experimental setup is similar to the one described in
Ref.~\cite{lamriben2010}, and is briefly recalled here
(Fig.~\ref{fig:schemaexp}a). Turbulence is generated by towing a
square grid at a velocity $V_g=1.0$~m~s$^{-1}$ from the bottom to
the top of a cubic glass tank, of side 60~cm, filled with 52~cm of
water. The grid consists in 8~mm thick bars with a mesh size
$M=40$~mm. The whole setup is mounted on a precision rotating
turntable of 2~m in diameter. Runs for three rotation rates,
$\Omega = 0.42, 0.84$ and 1.68~rad~s$^{-1}$ (4, 8 and 16 rpm), as
well as a reference run without rotation, have been carried out.
The initial Reynolds number based on the grid mesh is $Re_g = V_g
M / \nu = 40\,000$, and the initial Rossby number $Ro_g = V_g /
2\Omega M$ ranges from 7.4 to $30$, indicating that the flow in
the close wake of the grid is fully turbulent and weakly affected
by rotation. During the turbulence decay, the instantaneous Rossby
number, $Ro(t) = \langle {\bf u}^2 \rangle^{1/2} / 2 \Omega M$,
decreases with time down to $10^{-2}$, spanning a range in which
influence of rotation is expected. An important concern about grid
turbulence experiments in a confined rotating domain is the
excitation of reproducible inertial modes~\cite{bewley2007}. Here,
we use the modified grid introduced in Ref.~\cite{lamriben2010},
which was shown to significantly reduce the generation of these
modes. Consequently, turbulence can be considered here as almost
freely decaying and homogeneous,  a necessary condition for the
validity of the KHM equation (\ref{eq:vKH}).

Velocity measurements are performed in the rotating frame using a
corotating PIV system. Two velocity components $(u_x, u_z)$ are
measured, in a vertical $16 \times 16$~cm$^2$ field of view, where
$z$ is the rotation axis. During the decay of turbulence, $60$
image pairs are acquired by a double-frame $2048^2$ pixels camera,
at a rate of 1 pair per second. The PIV resolution, $1.3$~mm, is
sufficient to resolve the inertial range but fails to resolve the
dissipative scale (the Kolmogorov scale is of the order of 0.2~mm
right after the grid translation~\cite{morize2005}).

Only surrogates of $E({\bf r})$ and ${\bf F}({\bf r})$
(\ref{eq:defF}) can be computed from the measured 2D velocity
fields. These surrogate quantities are defined as
\begin{equation}\label{eq:defFt}
\widetilde{E}({\bf r}) = \langle \delta u_x^2 +\delta u_z^2
\rangle_{x,z}, \hspace{0.4cm} {\bf \widetilde F} ({\bf r}) =
\langle \delta {\bf u}(\delta u_x^2 + \delta u_z^2) \rangle_{x,z},
\end{equation}
where the spatial average is restricted to the measurement plane,
and ${\bf r} = r_x {\bf e}_x + r_z {\bf e}_z$. For each time after
the grid translation, these quantities are computed for all
separations ${\bf r}$ in the PIV field of view, and are
ensemble-averaged over 600 realizations of the turbulence decay.
The fields $\widetilde{E}({\bf r})$ and ${\bf \widetilde F}({\bf
r})$ are remapped on a spherical coordinate system $(r, \theta,
\phi)$, where $r = |{\bf r}|$, and $\theta$ is the polar angle
between ${\bf e}_z$ and ${\bf r}$;  the invariance with respect to
the (non-measured) azimuthal angle $\phi$ is assumed by
axisymmetry. Although relations between the surrogates
(\ref{eq:defFt}) and the exact 3-components quantities
(\ref{eq:defF}) can be derived for isotropic turbulence, no
general relation holds in the anisotropic case, so we do not apply
any correction weight in $\widetilde{E}$ and ${\bf
\widetilde{F}}$. Since only the surrogates are considered in this
paper, we simply drop the tildes $\widetilde{\cdot}$ in the
following.

The convergence of the statistics from experimental measurements
is very delicate to achieve, in particular for the computation of
${\bf F} ({\bf r})$, which is a 3rd order moment of a zero-mean
velocity increment. We found that, using a set of 600 realizations
of the turbulence decay, a convergence better than 5\% at small
scales, and of the order of 20\% at scales $r \simeq M$, could be
achieved for ${\bf F}({\bf r})$. The convergence for $E({\bf r})$
is better than 1\% for all scales up to $r \simeq M$.

\paragraph*{Energy distribution.-}

\begin{figure}
\centerline{\includegraphics[width=8cm]{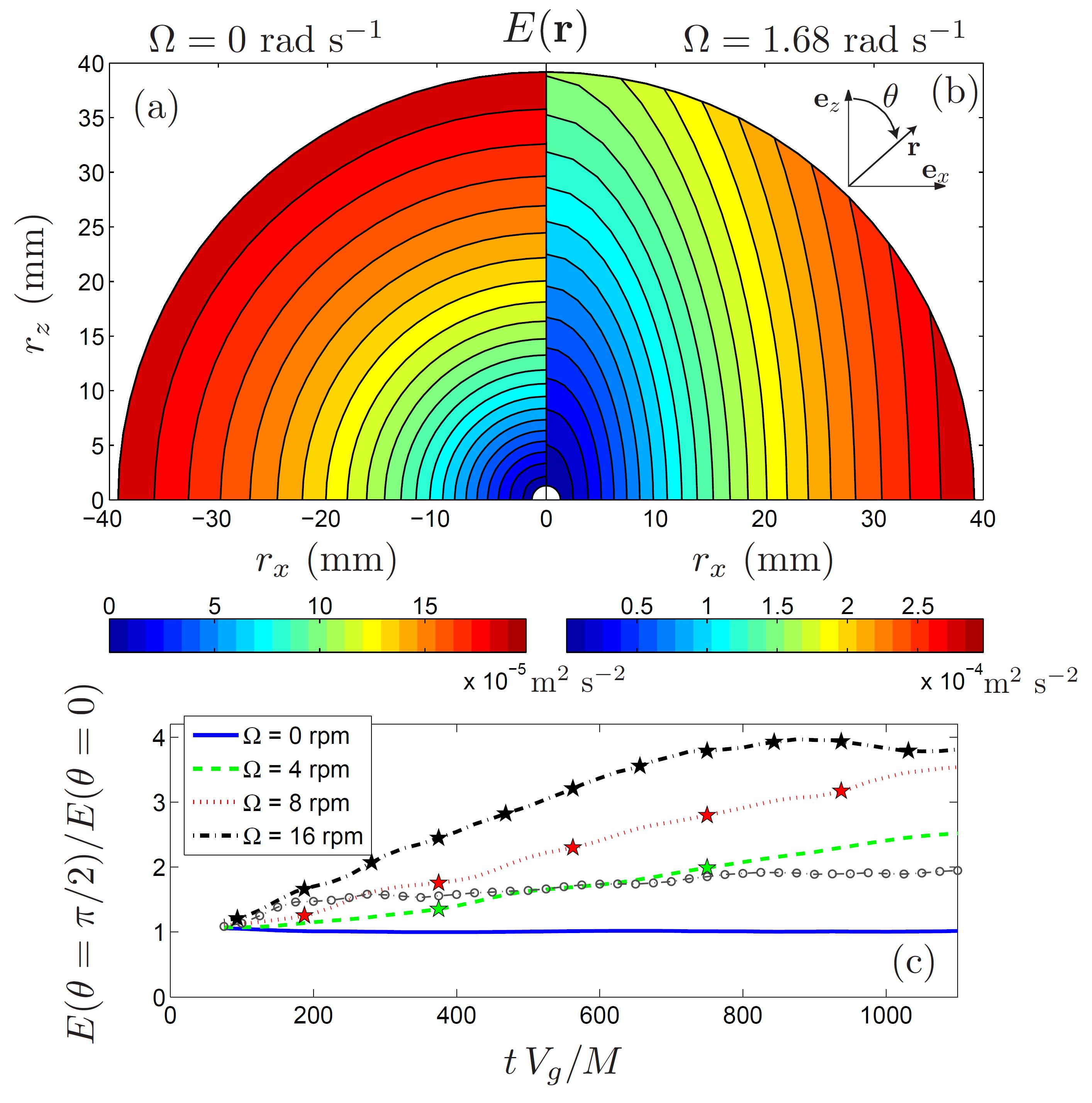}}
\caption{Energy distribution $E({\bf r})$ at time $t\,V_g/M=400$
after the grid translation, for (a) $\Omega=0$, and (b)
$\Omega=1.68$~rad~s$^{-1}$ (16 rpm). (c) Horizontal-to-vertical
energy ratio as a function of time at scale $r=10$~mm for various
$\Omega$; $\circ$: additional curve at $r=30$~mm for $\Omega =
16$~rpm. Stars indicate integer numbers of tank rotations.}
\label{fig:v2}
\end{figure}

The map of energy distribution $E({\bf r})$ for separations ${\bf
r}$ in the vertical plane is plotted in Fig. \ref{fig:v2}, at a
time $t\,V_g/M=400$ after the grid translation. The iso-$E$ curves
are found nearly circular for $\Omega = 0$ (Fig. \ref{fig:v2}a),
showing the good level of isotropy of our grid turbulence without
rotation. On the other hand, they are highly anisotropic at the
same time for $\Omega=16$~rpm (corresponding to 4.3 tank
rotations), with a strong depletion of $E({\bf r})$ along the
rotation axis $z$ (Fig. \ref{fig:v2}b). The depletion of $E({\bf
r})$ corresponds to an enhanced velocity correlation $R({\bf r})$
along the rotation axis, reflecting the classical trend towards a
2D flow invariant along $z$. Importantly, an isotropic energy
distribution is found in the 3 rotating cases just after the grid
translation, as demonstrated in Fig. \ref{fig:v2}(c) where the
time evolution of the horizontal-to-vertical energy ratio
$E(\theta=0)/E(\theta=\pi/2)$ is plotted for an inertial-range
separation $r=10$~mm. This confirms that the initial grid
turbulence is isotropic even when $\Omega \neq 0$, and that the
subsequent anisotropy growth is a pure effect of the background
rotation. Fig. \ref{fig:v2}(c) also shows that the anisotropy
growth rate is essentially proportional to
$\Omega$~\cite{staplehurst2008,moisy2011}. Interestingly, the
anisotropy is found more pronounced at small scales, as shown by
the lower anisotropy ratio plotted for $r = 30$~mm. It is worth
noting that this stronger anisotropy at small scales is in
contradiction with the naive assumption that large scales, having
a slower dynamics, are more affected by rotation than the faster
and supposedly still 3D small scales.

\begin{figure}
\centerline{\includegraphics[width=8cm]{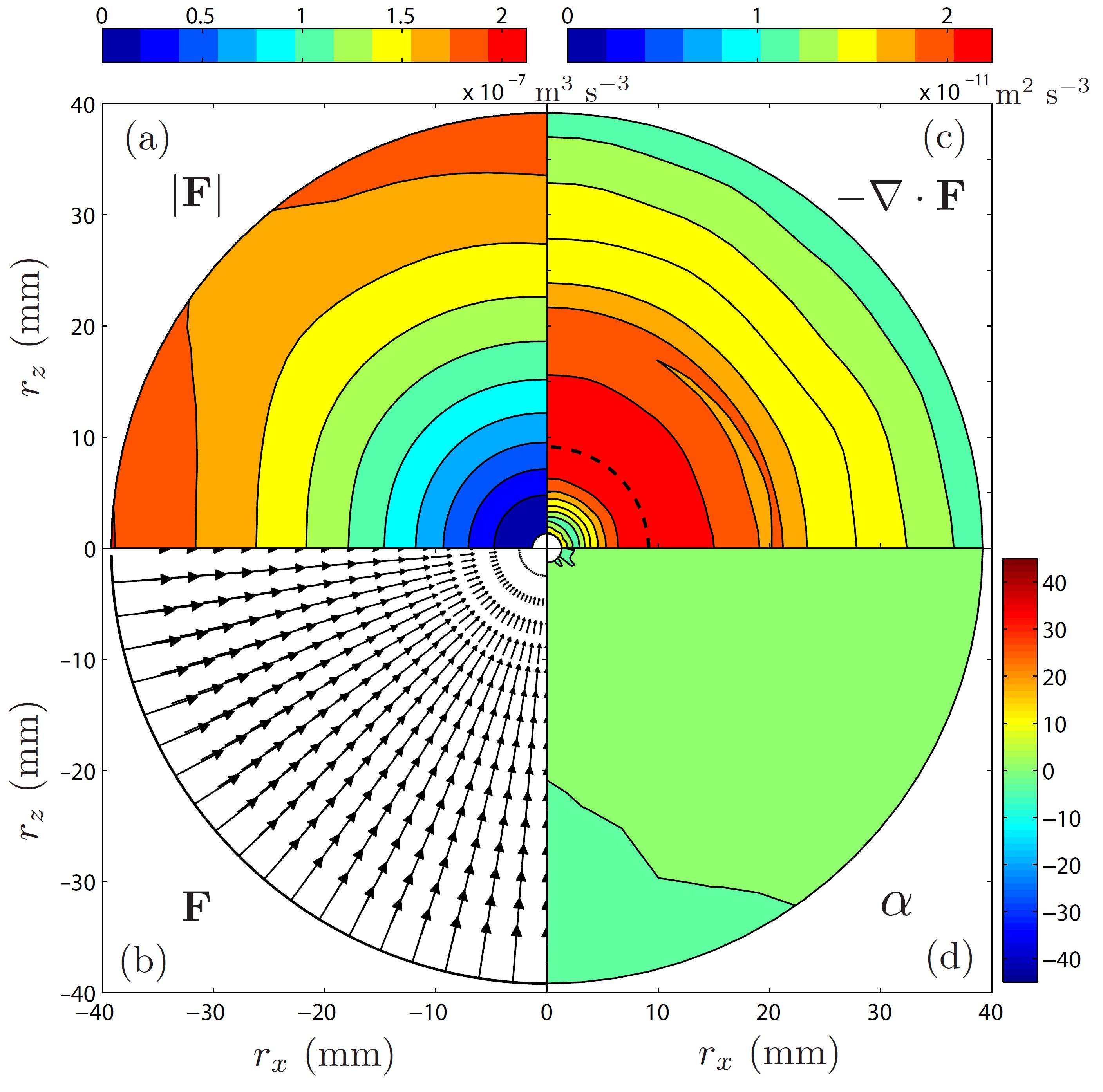}}
\caption{Energy flux density ${\bf F}({\bf r})$ in the
non-rotating case, at time $t\,V_g/M=400$ after the grid
translation. (a) Flux density magnitude $|{\bf F}|$. (b) Raw
vector field ${\bf F}$. (c) Energy flux $\nabla \cdot {\bf F}$.
(d) Deviation angle $\alpha ({\bf r})$, defined as $\sin
\alpha({\bf r}) = {\bf e}_y \cdot ({\bf e}_r \times {\bf F})/|{\bf
F}|$; iso-angle lines are separated by $5\degree$. The dashed line
in (c) shows the ``crest line'', following the local maximum of $-
\nabla \cdot {\bf F}$.}\label{fig:Fnonrot}
\end{figure}

\paragraph*{Energy transfers: isotropic case.-}
We now turn to the energy flux density, and we first present in
Fig.~\ref{fig:Fnonrot}(b) measurements of ${\bf F}({\bf r})$ for
$\Omega=0$, at the same time $t V_g / M = 400$. This vector field
is found remarkably radial, pointing towards the origin, giving
direct evidence of the isotropic energy cascade in the physical
space, from the large to the small scales,  in the non-rotating
case. Finer assessment of the isotropy of ${\bf F}$ can be
achieved by introducing the following three scalar quantities: the
deviation angle $\alpha({\bf r})$ from the radial direction
(Fig.~\ref{fig:Fnonrot}d), the magnitude $|{\bf F}|$
(Fig.~\ref{fig:Fnonrot}a), and the energy flux $\nabla \cdot {\bf
F}$ (Fig.~\ref{fig:Fnonrot}c). The very weak angle measured for $r
\leq M$, $\alpha({\bf r}) \simeq 2\degree \pm 2\degree$, confirms
the almost purely radial nature of ${\bf F}$. The isotropy of the
flux density magnitude is not as good: the iso-$|{\bf F}|$ are
nearly circular up to $r \simeq 30$~mm, but shows slight departure
from isotropy at larger $r$, suggesting that this quantity is very
sensitive to a residual anisotropy of the large-scale flow.
However, the iso-$\nabla \cdot {\bf F}$ remain remarkably circular
up to $r \simeq M$, showing that the residual large-scale
anisotropy has indeed a weak influence on the energy flux for $r
\leq M$. The energy flux $\nabla \cdot {\bf F}$ shows a broad
negative minimum in an annular region spanning over $r \simeq
5-20$~mm, providing an indication of the extent of the inertial
range (we recall that, in the inertial range,  $\nabla \cdot {\bf
F} = -4 \epsilon$), and decreases to zero both at small and large
scales.

\begin{figure}
\centerline{\includegraphics[width=8cm]{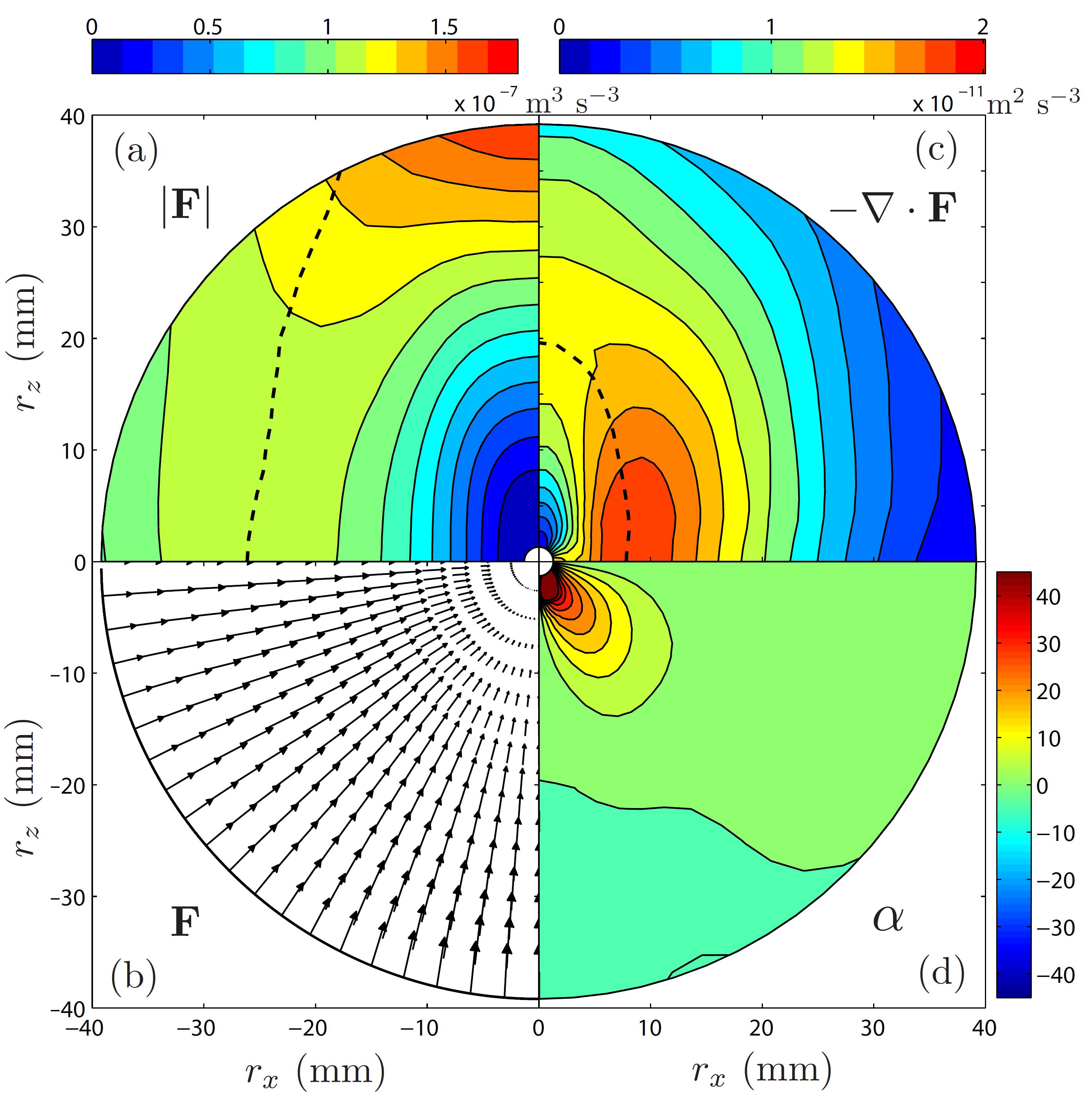}}
\caption{Energy flux density ${\bf F}({\bf r})$ in the rotating
case ($\Omega=16$~rpm), at time $t\,V_g/M=400$ after the grid
translation. Same layout as for Fig.
\ref{fig:Fnonrot}.}\label{fig:Frot}
\end{figure}

\paragraph*{Energy transfers: rotating case.-}
We consider now the energy transfers in the rotating case, shown
in Fig.~\ref{fig:Frot} at the same time $t\,V_g/M=400$.
Interestingly, the flux density ${\bf F}$ is found to remain
nearly radial for all separations, in qualitative agreement with
recent predictions \cite{galtier2009}, except at the smallest
scales, for $r < 10$~mm, where a marked deflection towards the
rotation axis is observed. Such horizontally tilted ${\bf F}$ is
indeed consistent with an asymptotic 2D flow, for which ${\bf F}$
must be a strictly horizontal vector, function of the horizontal
component of the separation only. This small-scale anisotropy is
best appreciated from the map of the deviation angle $\alpha$
(Fig.~\ref{fig:Frot}d), showing a region of nonzero $\alpha$ at
small scale only. Note that horizontally tilted ${\bf F}$ exists
only for intermediate angle $\theta$ since axisymmetry requires a
radial ${\bf F}$ for $\theta = 0$ and $\pi/2$. The 2D trend is
remarkably weak in terms of the orientation of ${\bf F}({\bf r})$
in the inertial range, compared to the strong anisotropy observed
for the energy distribution $E({\bf r})$ at comparable scales:
$\alpha$ is only in the range $0-10\degree$ in the inertial range,
and increases up to $25\degree \pm 5\degree$ for $r \rightarrow
0$, with no significant dependence with $\Omega$.

If we focus on the flux density magnitude $|{\bf F}|$, which is
essentially given by the radial component $- F_r = - {\bf F} \cdot
{\bf e}_r$, a clear anisotropy is now found  at all scales. This
suggests that the anisotropy of the energy transfers is mostly
driven by the $\theta$-dependence of $F_r$, and not by the growth
of a nonzero polar component $F_\theta = {\bf F} \cdot {\bf
e}_\theta$. The maximum of $|{\bf F}|$ is systematically
encountered near the rotation axis, at rather large scales,
centered around 50-80~mm (outside the range shown in
Fig.~\ref{fig:Frot}a). The local maximum of $|{\bf F}|$ on the
horizontal axis is encountered at smaller scales, as evidenced by
the crest line in Fig.~\ref{fig:Frot}(a).

\begin{figure}
\centerline{\includegraphics[width=8cm]{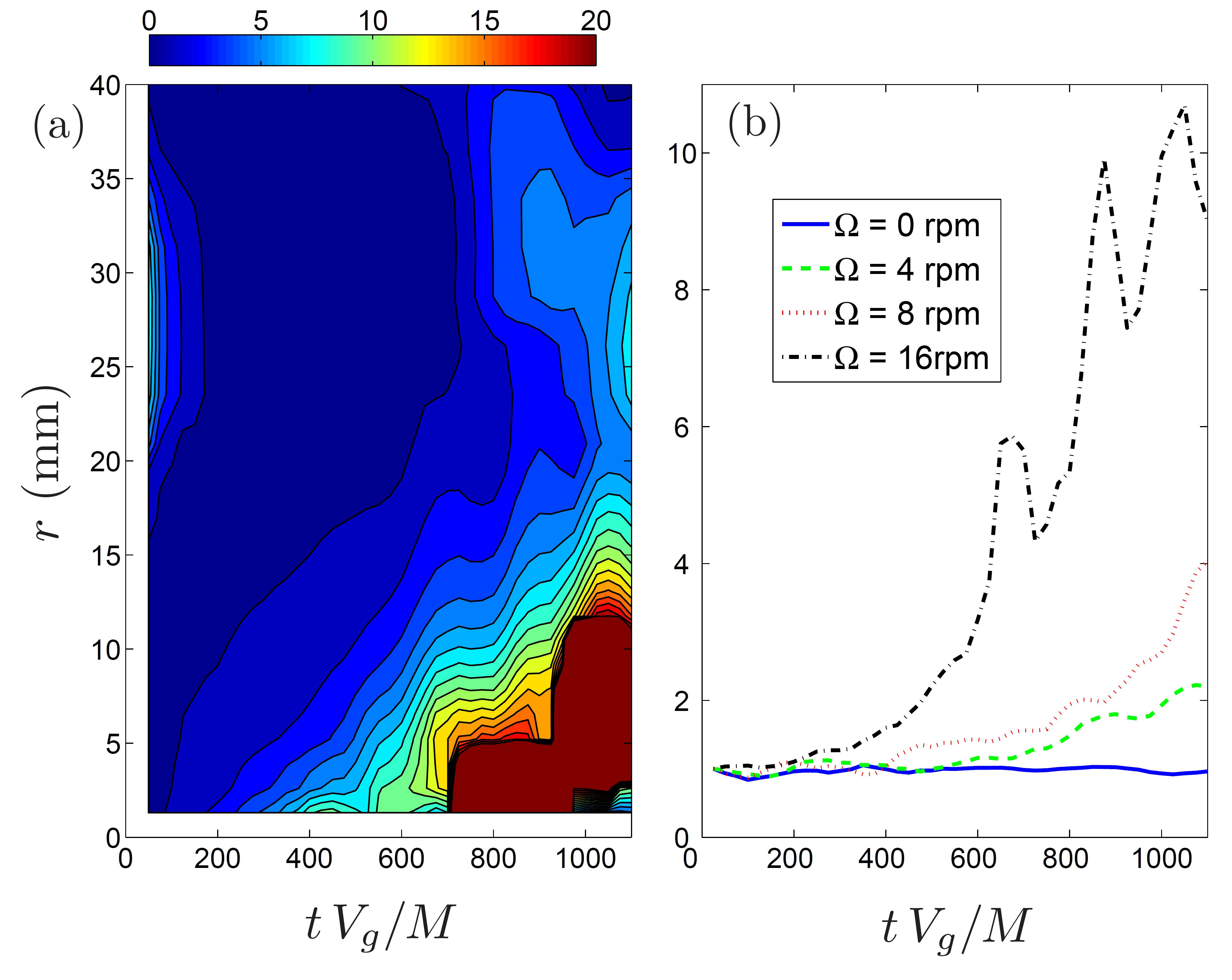}} \caption{(a)
Spatio-temporal diagram of the horizontal-to-vertical energy flux
ratio, $\nabla \cdot {\bf F}(\theta = \pi/2) / \nabla \cdot {\bf
F}(\theta = 0)$, showing the anisotropy growing from small to
large scales. (b) Time evolution of the energy flux ratio at scale
$r=10$~mm for various $\Omega$.} \label{fig:ratio}
\end{figure}

The flux map $\nabla \cdot {\bf F}$ (Fig.~\ref{fig:Frot}c) shows
an overall anisotropic structure similar to that of $|{\bf F}|$,
but essentially shifted towards smaller scales. The inertial
range, where the flux $\nabla \cdot {\bf F}$ is negative and
approximately constant, becomes vertically elongated as time
proceeds.   Actually, although $|{\bf F}|$ is maximum along the
rotation axis, it is spread over a wider range of scales, leading
to a  weaker flux $\nabla \cdot {\bf F}$ along $z$ than along $x$,
and hence a less intense vertical energy cascade. Here again, this
is consistent with a 2D trend, which should yield a vanishing
energy flux along the rotation axis. The  horizontal-to-vertical
flux ratio in Fig.~\ref{fig:ratio} illustrates this vanishing
vertical energy cascade as time proceeds, an effect which is
clearly enhanced as the rotation rate is increased.

It must be noted that the spatial structure of the flux $\nabla
\cdot {\bf F}$ is in good qualitative agreement with the KHM
equation (\ref{eq:vKH}). Indeed, neglecting the viscous term, the
vertically elongated region where $\nabla \cdot {\bf F}<0$ induces
a stronger  reduction of the velocity correlation $R$ along $x$
than along $z$, resulting in a relative growth of the vertical
correlation  along $z$, and hence a vertical depletion of the
energy distribution $E = 2(\langle {\bf u}^2 \rangle - R)$. We can
conclude that the measured flux density ${\bf F}$ contains,
through its divergence, a spatial structure consistent with the
anisotropy growth of $E$ observed in Fig~\ref{fig:v2}.
Interestingly, in line with the stronger anisotropy of $E({\bf
r})$ found at smaller scales, the flux is also found more
anisotropic at smaller scales. This is clearly demonstrated by the
spatio-temporal diagram in Fig.~\ref{fig:ratio}(a), showing that
the anisotropy first appears at small scales, and then propagates
towards larger scales as time proceeds.

\paragraph*{Conclusion.-}
We report the first direct measurements of the energy flux density
${\bf F}$ in the physical space in a decaying rotating turbulence
experiment. Although the alternative description of the energy
transfers in the spectral space is more natural for theory or
numerics~\cite{sagaut2008,cambon1989,waleffe1993,mininni2010}, the
direct use of the KHM equation (\ref{eq:vKH}) in the physical
space, which is better suited for experiments, reveals here new
and unexpected behaviors. The spatial structure of the measured
energy distribution and energy flux $\nabla \cdot {\bf F}$ are
found in good qualitative agreement with the KHM equation which,
to our knowledge, has never been assessed experimentally.
Surprisingly, the anisotropy growth of the energy distribution is
primarily driven by an almost radial, but orientation-dependent,
flux density ${\bf F}$, except at small scales where ${\bf F}$
shows a horizontal tilt, compatible with a trend towards a 2D
state. It is also demonstrated that the anisotropy is
paradoxically stronger at small scales, and propagates towards
larger scales as time proceeds, an unexpected result which should
motivate new theoretical efforts.

\begin{acknowledgments}

We acknowledge S. Galtier, J.-P. Hulin, and M. Rabaud for fruitful discussions,
and Triangle de la Physique for funding of the ``Gyroflow'' platform.

\end{acknowledgments}

\end{document}